\documentclass[aps,prd,preprint,preprintnumbers]{revtex4}
\usepackage{ae}
\usepackage{bm} 
\usepackage{color}
\usepackage{amssymb}
\usepackage{amsmath}
\usepackage{amsfonts}
\usepackage{graphicx}
\usepackage{slashed}
\usepackage{setspace}
\usepackage{subfigure}
\usepackage{ulem}
\usepackage{bbold}

\newcommand{\Eqref}[1]{Eq.~\eqref{#1}}

\allowdisplaybreaks

\begin{document}

\title{Large $N$ external-field quantum electrodynamics}

\author{Felix Karbstein}\email{felix.karbstein@uni-jena.de}
\affiliation{Helmholtz-Institut Jena, Fr\"obelstieg 3, 07743 Jena, Germany}
\affiliation{GSI Helmholtzzentrum f\"ur Schwerionenforschung, Planckstra\ss e 1, 64291 Darmstadt}
\affiliation{Theoretisch-Physikalisches Institut, Abbe Center of Photonics, \\ Friedrich-Schiller-Universit\"at Jena, Max-Wien-Platz 1, 07743 Jena, Germany}

\date{\today}

\begin{abstract}
We advocate the study of external-field quantum electrodynamics with $N$ charged particle flavors.
Our main focus is on the Heisenberg-Euler effective action for this theory in the large $N$ limit which receives contributions from all loop orders.
The contributions beyond one loop stem from one-particle reducible diagrams. We show that
specifically in constant electromagnetic fields the latter are generated by the one-loop Heisenberg-Euler effective Lagrangian.
Hence, in this case the large $N$ Heisenberg-Euler effective action can be determined explicitly at any desired loop order. We demonstrate that further analytical insights are possible for electric-and magnetic-like field configurations characterized by the vanishing of one of the secular invariants of the electromagnetic field and work out the all-orders strong field limit of the theory.
\end{abstract}

\maketitle

\subsection{Introduction}\label{sec:intro_N}

The present article is devoted to the study of quantum electrodynamics (QED) with $N$ charged particle flavors of equal charge $e$ and mass $m$ in the presence of a prescribed external electromagnetic field $\bar A$.
This theory resembles standard QED but features $N$ generations of electrons and positrons coupling to the electromagnetic field via the charge $e$.
Throughout this work we choose this coupling such that the combination $Ne^2$ does not change with $N$. This choice implies $e\sim N^{-1/2}$.
To arrive at a meaningful deformation of standard external-field QED we moreover demand $\bar A\sim N^{1/2}$, which guarantees that $e\bar A\sim N^0$  \cite{Karbstein:2019wmj}.
Without this additional assumption, the considered theory would generically be dominated by physics at zero field because $e\bar A\sim N^{-1}\to 0$ for sufficiently large values of $N\gg1$.

We are convinced that this generalization of standard external-field QED constitutes an interesting deformation of the original theory. As detailed in Sec.~\ref{sec:foundations_N} below, the flavor number $N$ in particular provides a new parameter allowing to organize Feynman diagrams contributing to a given quantity by classifying them in terms of their scaling with $N$.
The main focus of the present work is on the emergence and structure of the Heisenberg-Euler effective action for this theory in the large $N$ limit, or more specifically the 't Hooft limit $N\to\infty$, $Ne^2$ fixed \cite{tHooft:1973alw}.
In this context, we lay the foundations for future non-perturbative study of the large $N$ Heisenberg-Euler effective action in arbitrarily strong constant electromagnetic fields, and work out the all-orders strong field limit of the theory in the parameter regime where a perturbative loop expansion is possible.

Our article is organized as follows: In Sec.~\ref{sec:foundations_N} we briefly recall the foundations of external-field QED with $N$ charged particle flavors. Subsequently, in Sec.~\ref{sec:HE_N} we derive the Heisenberg-Euler effective action for this theory in the large $N$ limit. After working out its structure in generic external fields in Sec.~\ref{sec:generic_N}, in Sec.~\ref{sec:const_N} we focus on constant external fields where additional analytical insights are possible. We in particular show that in this case the effective action for $N\to\infty$ is fully determined by the one-loop Heisenberg-Euler effective Lagrangian. Finally, we specialize to magnetic- and electric-like field configurations characterized by the vanishing of one of the secular invariants of the electromagnetic field. In this context, we provide an alternative derivation of the all-loop strong-field limit first derived in Ref.~\cite{Karbstein:2019wmj}. Finally we end with conclusions and an outlook in Sec.~\ref{sec:concls_N}.

\section{Foundations}\label{sec:foundations_N}

The microscopic Lagrangian of external-field QED with $N$ fermion generations considered throughout this work can be cast into the form,
\begin{equation}
{\cal L}(\bar\psi,\psi,q,\bar A)	= \bar\psi\bigl({\rm i}\slashed{D}[\bar A + q]-m\bigr)\psi -\frac{1}{4}\bar F_{\mu\nu}\bar F^{\mu\nu}-\frac{1}{4}Q_{\mu\nu}Q^{\mu\nu}\,, \label{eq:sfQEDlargeN}
\end{equation}
with gauge covariant derivative  $D_\mu[A]=\partial_\mu-{\rm i}eA_\mu$, $\slashed{D}=\gamma^\mu D_\mu$ and gamma matrices $\gamma^\mu$; see Refs.~\cite{Dittrich:1985yb,Fradkin:1991zq,Dittrich:2000zu,Dunne:2004nc,Marklund:2006my,DiPiazza:2011tq,Battesti:2012hf,King:2015tba,Karbstein:2016hlj,Gies:2016yaa,Karbstein:2019oej} for pertinent reviews of standard external-field QED. Here, the scalings $e^2\sim N^{-1}$ and $e\bar F\sim e\bar A\sim N^0$ and the
notation $\bar\psi(\cdot)\psi=\sum_{i=1}^N\bar\psi^{(i)}(\cdot)\psi^{(i)}$, where $i$ labels the fermion flavors,
are implicitly understood in \Eqref{eq:sfQEDlargeN}.
Moreover, in the present context $\psi^{(i)}$ and $q$ denote quantized spinor and gauge fields, and $\bar A$ is the four-potential of the prescribed non-quantized external field. The associated field strength tensors are $\bar F^{\mu\nu}=\partial^\mu\bar A^\nu-\partial^\nu\bar A^\mu$ and $Q^{\mu\nu}=\partial^\mu q^\nu-\partial^\nu q^\mu$.
For $N=1$ and $\bar A\to 0$ the Lagrangian of standard QED at zero field is recovered.

The Feynman rules of $N$ flavor external-field QED resemble those of standard QED. As conventionally done in studies of standard QED subjected to a prescribed classical field, the action of the external field $\bar A$ on the dynamics of the theory can be fully encoded in dressed fermion propagators \cite{Furry:1951zz}. The resulting fermion propagator accounts for the dressing of the fermion line to all orders in the classical field $e\bar A\sim N^0$.
At the same time, each fermion loop comes with an overall factor of $N$ as quantum fluctuations generically encompass all degenerate fermion flavors.
On the other hand a coupling to the quantum photon field is mediated by the elementary charge and thus comes with a factor of $e\sim N^{-1/2}$. This directly implies a factor of $N^{-1}$ for each internal photon line.

In turn, the one-loop Heisenberg-Euler effective action  \cite{Heisenberg:1935qt,Weisskopf:1996bu,Schwinger:1951nm} scales as $\Gamma_{\rm HE}^{1\text{-loop}}\sim N$, and all higher-loop one-particle irreducible (1PI) contributions are parametrically suppressed relatively to this contribution by at least one power of $N^{-1}$.
As another important example consider the physical photon propagator: it can be straightforwardly inferred that the one-loop bubble resummed photon propagator scales as $N^0$. At the same time, the contribution featuring an additional internal photon line is relatively suppressed with a factor of $N^{-1}$.

A distinct difference between external-field and standard zero-field QED is the emergence of finite physical tadpole contributions: due to the dressing in the external field, e.g., a tadpole formed by a fermion loop contracted with a single photon line constitutes a viable, generically non-vanishing building block to a Feynman diagram describing a given process. 
Upon connection to another fermion line this particular tadpole contribution scales as $NN^{-1}=N^0$, and thus can be attached to any given diagram without changing its scaling with $N$. Precisely this property and the fact that they even persist to contribute in homogeneous constant electromagnetic fields~\cite{Gies:2016yaa} makes such tadpole contributions constituent building blocks of the Heisenberg-Euler effective action in the large $N$ limit; cf.~Sec.~\ref{sec:HE_N} below. Note, however, that all physical tadpole contributions vanish for the special cases of constant crossed and plane-wave fields \cite{Gies:2016yaa,Ahmadiniaz:2019nhk}.

Interestingly this implies that in $N$-flavor external-field QED the all-order resummed bubble-type polarization corrections to the electron mass operator evaluated in Ref.~\cite{Mironov:2020gbi} constitute the full result of the electron mass operator in constant crossed fields at order $N^0$.
On the other hand, upon summation over $\ell$, the $N$-flavor versions of the irreducible $\ell$-loop diagrams giving the dominant strong-field behavior of the 1PI part of the $\ell$-loop Euler-Heisenberg effective Lagrangian discussed in Ref.~\cite{Dunne:2021acr} constitute the full result for the 1PI part of the Heisenberg-Euler effective action in generic constant fields at order $N^0$.

\section{Heisenberg-Euler effective action}\label{sec:HE_N}

A central quantity in
the study of the effective nonlinear interactions of macroscopic electromagnetic fields in the QED vacuum is the Heisenberg-Euler effective action \cite{Heisenberg:1935qt,Weisskopf:1996bu,Schwinger:1951nm} mentioned already in the previous section.

\subsection{Generic external fields}\label{sec:generic_N}

The Heisenberg-Euler effective action $\Gamma_\text{HE}[\bar A]$ follows from the Lagrangian~\eqref{eq:sfQEDlargeN} upon integrating out the quantum fields as \cite{Dittrich:1985yb,Gies:2016yaa}
\begin{equation}
	{\rm e}^{{\rm i}\Gamma_\text{HE}[\bar A]}=\int{\cal D}q\int{\cal D}\bar\psi\int{\cal D}\psi\,{\rm e}^{{\rm i}\int{\cal L}(\bar\psi,\psi,q,\bar A)}\,. \label{eq:exp(iGammaHElargeN)}
\end{equation}
To keep notations compact, throughout this work we employ the shorthand notations $\int_x\equiv\int{\rm d}^4x$ and $\int_k\equiv\int{\rm d}^4k/(2\pi)^4$ for integrations over position and momentum space, respectively. Besides, we use $\int$ if the integration can be performed in position or momentum space.
Shifting $q_\nu\to q_\nu+e\int\bar\psi\gamma^\mu\psi D_{\mu\nu}:={\cal J}_\nu$, \Eqref{eq:exp(iGammaHElargeN)} can be rewritten as
\begin{equation}
{\rm e}^{{\rm i}\Gamma_\text{HE}[\bar A]}={\rm e}^{-{\rm i}\frac{1}{4}\int\bar F_{\mu\nu}\bar F^{\mu\nu}}\int{\cal D}\bar\psi\int{\cal D}\psi \int{\cal D}q\ {\rm e}^{{\rm i}\int \bar\psi({\rm i}\slashed{D}[\bar A+{\cal J}]-m)\psi-{\rm i} \frac{1}{2}\iint{\cal J}_\mu(D^{-1})^{\mu\nu}{\cal J}_\nu}\,.
\label{eq:exp(iGammaHElargeN2)}
\end{equation}
Here, $D^{\mu\nu}$ denotes the photon propagator and $(D^{-1})^{\mu\nu}$ its inverse.
In momentum space and accounting for a gauge fixing term we have
\begin{equation}
 D^{\mu\nu}(p,p')=(2\pi)^4\delta(p+p')\,\frac{1}{p^2-{\rm i}\epsilon}\Bigl(g^{\mu\nu}-(1-\xi)\frac{p^\mu p^\nu}{p^2-{\rm i}\epsilon}\Bigr)\,,
\end{equation}
with $\xi=1$ in Feynman gauge.
Note that \Eqref{eq:exp(iGammaHElargeN2)} with $\int{\cal D}q\to\int{\cal DJ}$ can also be understood as a Hubbard-Stratonovich transformation \cite{Stratonovich:1957,Hubbard:1959ub} of the non-local four-fermion theory obtained by carrying out the Gaussian integral over $q$ in \Eqref{eq:exp(iGammaHElargeN)}.

Equation~\eqref{eq:exp(iGammaHElargeN2)} is particularly suited for studying external-field QED in the large $N$ limit. Being interested only in the leading contribution to $\Gamma_{\rm HE}[\bar A]$ for $N\to\infty$, we can completely drop the integration over the quantum photon field $q$, and simply identify ${\cal J}$ with its expectation value $\bar{j}_\nu:=\langle{\cal J}_\nu\rangle=e\int\langle\bar\psi\gamma^\mu\psi\rangle D_{\mu\nu}$.
Diagrammatically, this means that Feynman diagrams with internal photon lines are not taken into account.
Higher orders in an expansion of the fluctuation field $q$ are suppressed parametrically by powers of $1/N$: the associated Feynman diagrams contain at least one internal photon line.

In turn, we arrive at
\begin{equation}
{\rm e}^{{\rm i}\Gamma_\text{HE}[\bar A]} \big|_{N\to\infty}={\rm e}^{-{\rm i}\frac{1}{4}\int\bar F_{\mu\nu}\bar F^{\mu\nu}}\int{\cal D}\bar\psi\int{\cal D}\psi \ {\rm e}^{{\rm i}\int \bar\psi({\rm i}\slashed{D}[\bar A+\bar j]-m)\psi-{\rm i} \frac{1}{2}\iint\bar j_\mu(D^{-1})^{\mu\nu}\bar j_\nu}\,.
\label{eq:exp(iGammaHElargeN3)}
\end{equation}
The exponential in \Eqref{eq:exp(iGammaHElargeN3)} is bilinear in the quantum field, such that the functional integrations over the quantum fields $\bar\psi$ and $\psi$ can be performed exactly, yielding
 \begin{align}
	\int{\cal D}\bar\psi\int{\cal D}\psi\ {\rm e}^{{\rm i}\int \bar\psi({\rm i}\slashed{D}[A]-m)\psi}={\det}^N\bigl(-{\rm i}\slashed{D}[A]+m\bigr)=:{\rm e}^{{\rm i}S_\psi[A]}\,.  \label{eq:Spsi}
\end{align}
Correspondingly, we have
\begin{equation}
 \Gamma_{\rm HE}[\bar A]\big|_{N\to\infty}=-\frac{1}{4}\int\bar F_{\mu\nu}\bar F^{\mu\nu}+S_\psi[\bar A+\bar j]-\frac{1}{2}
\iint \bar{j}_\mu(D^{-1})^{\mu\nu}\bar{j}_\nu\,. \label{eq:largeNGamma}
\end{equation}
Note that the second term in \Eqref{eq:largeNGamma} amounts to the expression for the $N$-flavor one-loop Heisenberg-Euler effective action evaluated in the prescribed field $\bar A+\bar j$, i.e., $\Gamma_\text{HE}^{1\text{-loop}}[\bar A+\bar j]=S_\psi[\bar A+\bar j]$. The latter follows from the standard one-loop Heisenberg-Euler effective action upon multiplication with an overall factor of $N$; cf. \Eqref{eq:Spsi}.

Determining the equations of motion of the classical field $\bar{j}$ from \Eqref{eq:largeNGamma} by variation for $\bar j$, we find
\begin{equation}
\frac{\delta}{\delta \bar{j}_\mu}	\Gamma_{\rm HE}[\bar A]\big|_{N\to\infty}=e\langle\bar\psi\gamma^\mu\psi\rangle-\int(D^{-1})^{\mu\nu}\bar{j}_\nu=0 \quad\leftrightarrow\quad \bar{j}_\nu=e\int\langle\bar\psi\gamma^\mu\psi\rangle D_{\mu\nu}\,,
\end{equation}
in line with the identification employed above \Eqref{eq:exp(iGammaHElargeN2)}.
We emphasize that this expectation value is to be evaluated self-consistently in the presence of both the external field $\bar A$ and the current $\bar{j}$ itself.

Noting that the inverse of the Dirac propagator in this background is given by $G^{-1}[\bar A+\bar{j}]=-{\rm i}\slashed{D}[\bar A+\bar{j}]+m$, one can straightforwardly show that 
$\langle\bar\psi\gamma^\mu\psi\rangle={\rm i}N{\rm Tr}\{G[\bar A+\bar{j}]\gamma^\mu\}$. The trace ${\rm Tr}\{\cdot\}={\rm tr}_\gamma{\rm tr}_x\{\cdot\} ={\rm tr}_\gamma{\rm tr}_p\{\cdot\} $ is over both spinor indices and coordinate or momentum space, respectively.
This immediately implies that
\begin{align}
\bar{j}_\mu&={\rm i}Ne\int D_{\mu\nu}{\rm Tr}\bigl\{\gamma^\nu G[\bar A+\bar{j}]\bigr\}\nonumber\\
&= {\rm i}Ne\int D_{\mu\nu}\sum_{n=0}^\infty{\rm Tr}\bigl\{\gamma^\nu G[\bar A]\bigl(e\bar{\slashed{j}}G[\bar A]\bigr)^n\bigr\}\,, \label{eq:currentlargeN}
\end{align}
where we employed an expansion in the number of current insertions in the last step.
Analogously, we have
\begin{align}
 S_\psi[\bar A+\bar j]
 &= -{\rm i}N\ln\det\bigl(G^{-1}[\bar A+\bar{j}]\bigr) \nonumber\\
 &= \underbrace{-{\rm i}N\ln\det\bigl(G^{-1}[\bar A]\bigr)}_{=S_\psi[\bar A]}
 +{\rm i}N\sum_{n=1}^\infty\frac{1}{n}{\rm Tr}\bigl\{\bigl(e\bar{\slashed{j}}G[\bar A]\bigr)^n\bigr\}\,.\label{eq:lndetlargeN}
\end{align}
To arrive at the result in the second line of \Eqref{eq:lndetlargeN}, we made use of the identity $\ln\det(\cdot)={\rm Tr}\ln(\cdot)$.

On the other hand, the all-order Taylor expansion of $S_\psi[\bar A+\bar j]$ can be cast into the form
\begin{equation}
	S_\psi[\bar A+\bar j]=S_\psi[\bar A]+\sum_{n=1}^\infty\frac{1}{n!}\int\ldots\int \bigl(S^{(n)}_\psi[\bar A]\bigr)^{\sigma_1\ldots\sigma_n}\bar j_{\sigma_1}\ldots \bar j_{\sigma_n}\,,
	\label{eq:SpsiTaylor}
\end{equation}
with
\begin{equation}
 \big(S_\psi^{(n)}[\bar A]\big)^{\sigma_1\ldots\sigma_n}:= \frac{\delta^n S_\psi[A]}{\delta A_{\sigma_1}\ldots\delta A_{\sigma_n}}\bigg|_{A=\bar A}\,. \label{eq:Spsi^n}
\end{equation}
The $n$th contribution to \Eqref{eq:SpsiTaylor} involves $n$ integrals: each contraction of a pair of Minkowski indices comes with an integration over the associated common variable.

A comparison of Eqs.~\eqref{eq:lndetlargeN} and \eqref{eq:SpsiTaylor} allows us to infer the following identity,
\begin{equation}
 {\rm tr}_\gamma\bigl\{e\gamma^{\sigma_1}G[\bar A]\ldots e\gamma^{\sigma_n}G[\bar A]\bigr\}=\frac{1}{{\rm i}N}\frac{1}{(n-1)!}\bigl(S^{(n)}_\psi[\bar A]\bigr)^{\sigma_1\ldots\sigma_n}
 \label{eq:traceloopngammas}
\end{equation}
for $n\in\mathbb{N}$.
With the help of this identity the current~\eqref{eq:currentlargeN} can be represented as
\begin{align}
\bar{j}_\mu=\int D_{\mu\sigma}\sum_{n=0}^\infty \frac{1}{n!}\int\ldots\int \bigl(S^{(n+1)}_\psi[\bar A]\bigr)^{\sigma\sigma_1\ldots\sigma_n} \,\bar{j}_{\sigma_1}\ldots \bar{j}_{\sigma_n} \,,
\label{eq:currentlargeNexpl}
\end{align}
and 
explicitly isolating the contribution linear in $\bar{j}$ from the sum in \Eqref{eq:currentlargeNexpl} and combining it with the term on its left-hand side as
\begin{align}
\bar{j}_{\mu} = \int\bigl[\bigl(D^{-1}-S^{(2)}_\psi[\bar A]\bigr)^{-1}\bigr]_{\mu\sigma}\sum_{n\in\mathbb{N}_0\setminus\{1\}} \frac{1}{n!}\int\ldots\int \bigl(S^{(n+1)}_\psi[\bar A]\bigr)^{\sigma\sigma_1\ldots\sigma_n} \,\bar{j}_{\sigma_1}\ldots \bar{j}_{\sigma_n} \,. \label{eq:currentlargeNexplfinal}
\end{align}
Moreover, upon insertion of \Eqref{eq:SpsiTaylor}, \Eqref{eq:largeNGamma} becomes
\begin{align}
 \Gamma_{\rm HE}[\bar A]\big|_{N\to\infty}=&-\frac{1}{4}\int\bar F_{\mu\nu}\bar F^{\mu\nu} + S_\psi[\bar A]
 \nonumber\\
 &+\sum_{n\in\mathbb{N}\setminus\{2\}}^\infty\frac{1}{n!}\int\ldots\int \bigl(S^{(n)}_\psi[\bar A]\bigr)^{\sigma_1\ldots\sigma_n} \, \bar{j}_{\sigma_1}\ldots \bar{j}_{\sigma_n} \nonumber\\
 & -\frac{1}{2}
\iint \bar{j}_\mu\bigl(D^{-1}-S^{(2)}_\psi[\bar A]\bigr)^{\mu\nu}\bar{j}_\nu\,.
 \label{eq:largeNGammafinal}
\end{align}

Equations~\eqref{eq:currentlargeNexplfinal} and \eqref{eq:largeNGammafinal} imply that all the non-trivial contributions to $\Gamma_\text{HE}[\bar A]|_{N\to\infty}$ can be expressed in terms of integrals contracting effective photon vertices $S_\psi^{(n)}[\bar A]$ generated by functional derivatives of the one-loop effective action $S_\psi[A]$.

\subsection{Constant external fields}\label{sec:const_N}

In constant electromagnetic fields the situation is particularly simple.
The reason for this is that constant fields cannot transfer momentum to virtual charged particle loops.
This and the demand of gauge invariance immediately implies that in momentum space the current $\bar j$ can be cast into the form \cite{Karbstein:2017gsb}
\begin{equation}
 \bar j_\mu(p)=\frac{1}{p^2-{\rm i}\epsilon}(2\pi)^4\delta(p)\Bigl[c(\bar F)(p\bar F)_\mu+c_*(\bar F)(p{}^*\!\bar F)_\mu+{\cal O}(p^3)\Bigr], \label{eq:barjconst}
\end{equation}
with {\it a priori} undetermined constant scalar coefficients $c(\bar F)$ and $c_*(\bar F)$ encoding the all-order dependences on both the background field $\bar F$ and the fine-structure constant $\alpha=e^2/(4\pi)$.

In turn, the large-$N$ all-orders non-perturbative Heisenberg-Euler effective Lagrangian in constant fields ${\cal L}_{\rm HE}|_{N\to\infty}=\Gamma_{\rm HE}|_{N\to\infty}/V^{(4)}$, with space-time volume $V^{(4)}=\int{\rm d}^4x$, could be determined by plugging \Eqref{eq:barjconst} into \Eqref{eq:largeNGamma} and minimizing the latter expression for the constant coefficients $c$ and $c_*$ at given $\bar F$. For inhomogeneous fields the structure of $\bar j$ is much less constrained, and such a minimization would be way more challenging.

However, in the present work we anyhow make use of perturbative expansions in the number of current insertions and proceed working with \Eqref{eq:largeNGammafinal}.
To this end, it is instructive to note that in constant external fields, the effective couplings~\eqref{eq:Spsi^n} in momentum space can be expressed as \cite{Gies:2016yaa}
\begin{equation}
 \bigl(S^{(n)}_\psi[\bar A]\bigr)^{\sigma_1\ldots\sigma_n}(p_1,\ldots, p_n)  =(2{\rm i})^n (2\pi)^4\,\delta\Bigl(\sum_{j=1}^n p_j\Bigr)\prod_{j=1}^n\biggl(p_j^{\mu_j} g^{\nu_j\sigma_j}
 \frac{\partial}{\partial\bar F^{\mu_j\nu_j}}\biggr) {\cal L}^{1\text{-loop}}_\text{HE}\bigl({\cal F},{\cal G}^2\bigr) \label{eq:Sn_constfield}
\end{equation}
and $S_\psi[\bar A]= \Gamma^{1\text{-loop}}_\text{HE}\bigl({\cal F},{\cal G}^2\bigr)$.
Hence, in this case an explicit evaluation of \Eqref{eq:largeNGammafinal} requires only input from the $N$-flavor constant-field Heisenberg-Euler Lagrangian at one loop ${\cal L}^{1\text{-loop}}_\text{HE}$.
Recall, that for $\bar F=\text{const}.$ the effective Lagrangian ${\cal L}_\text{HE}$ depends on the external field only via the gauge and Lorentz invariant scalars ${\cal F}=\frac{1}{4}\bar F_{\mu\nu}\bar F^{\mu\nu}=\frac{1}{2}(\vec{B}^2-\vec{E}^2)$ and ${\cal G}^2$, with pseudo-scalar ${\cal G}=\frac{1}{4}\bar F_{\mu\nu} {}^\star\!\bar F^{\mu\nu}=-\vec{B}\cdot\vec{E} $.

Upon plugging \Eqref{eq:Sn_constfield} into Eqs.~\eqref{eq:currentlargeNexplfinal} and \eqref{eq:largeNGammafinal} the constant-field Heisenberg-Euler effective Lagrangian in the 't Hooft limit can be readily worked out at any desired loop-order. 
Subsequently, we work with the on-shell renormalized result for ${\cal L}^{1\text{-loop}}_\text{HE}$, implying that also the resulting expression for $ \Gamma_{\rm HE}[\bar A]\big|_{N\to\infty}$ is on-shell renormalized.
Because ${\cal L}^{1\text{-loop}}_\text{HE}|_{\bar F=\text{const.}}$ can be expressed in terms of a single parameter integral over propertime $s$, generically the explicit representation of the $\ell$-loop contribution involves $\ell$ parameter integrals over propertime parameters $s_1,\ldots,s_\ell$.

\pagebreak

In a next step we focus on the special case of a constant electromagnetic field characterized by ${\cal G}=0$ and only ${\cal F}\neq0$. Here, our main interest is on the strong field limit of $\Gamma_\text{HE}({\cal F})|_{N\to\infty}$. We demonstrate that further analytical insights are possible in this specific case, and provide an alternative derivation of the all-loop strong field limit first worked out by Ref.~\cite{Karbstein:2019wmj}.

Particularly to identify the relevant contributions to \Eqref{eq:largeNGammafinal} which need to be accounted for a reliable extraction of the strong field limit we make use of the results of Ref.~\cite{Karbstein:2017gsb}, which showed that any possible 1PR diagram in constant fields can be generated from lower loop order diagrams by repeated action of the operator $(\partial{\cal L}_{1{\rm PI}}/\partial\bar F^{\mu\nu})\partial/\partial\bar F_{\mu\nu}$.
This operator adds an additional 1PI tadpole to a given lower loop diagram by trading a coupling to the external field $\bar F$ for a 1PI tadpole. In this way, both 1PI and 1PR tadpole structures are induced; the latter emerge from the dressing of 1PI tadpoles with additional 1PI tadpoles. In the limit of $N\to\infty$ we have ${\cal L}_{1{\rm PI}}= {\cal L}^{1\text{-loop}}_\text{HE}$.
Aiming at the construction of 1PR contributions to $\Gamma_{\rm HE}$, the initial quantity seeding all these contributions is again given by ${\cal L}^{1\text{-loop}}_\text{HE}$.
Correspondingly, the number of derivatives for $\bar F$ acting on a given factor of ${\cal L}^{1\text{-loop}}_\text{HE} $ counts the number of couplings of this particular 1PI contribution to other 1PI building blocks.
Thereby, a factor of $(\partial/\partial\bar F)^n {\cal L}^{1\text{-loop}}_\text{HE}$ constitutes an effective coupling for $n$ tadpoles.

For completeness, we note that when aiming at the determination of a specific 1PR diagram along these lines, it may be convenient to momentarily introduce labeled background fields dressing particular fermion lines in lower loop diagrams during the construction process. The labels are to be removed after all derivatives are taken.
This {\it trick} allows to explicitly choose the particular fermion line to which the additional 1PI tadpole is going to be attached by the application of the above operator.

As in constant fields ${\cal L}^{1\text{-loop}}_\text{HE}={\cal L}^{1\text{-loop}}_\text{HE}({\cal F},{\cal G}^2)$ we have
\begin{equation}
 \frac{\partial {\cal L}^{1\text{-loop}}_\text{HE}}{\partial\bar F^{\mu\nu}}\frac{\partial}{\partial \bar F_{\mu\nu}} =
 \frac{1}{2}\biggl(\bar F_{\mu\nu} \frac{\partial {\cal L}^{1\text{-loop}}_\text{HE}}{\partial{\cal F}}  + 2\,{}^*\!\bar F_{\mu\nu}\, {\cal G}\frac{\partial {\cal L}^{1\text{-loop}}_\text{HE}}{\partial{\cal G}^2}\biggr) \frac{\partial}{\partial \bar F_{\mu\nu}} \,. \label{eq:dFd}
\end{equation}
Using that ${\cal L}^{1\text{-loop}}_\text{HE}$ and thus also $\partial {\cal L}^{1\text{-loop}}_\text{HE}/\partial{\cal G}^2$ is regular for ${\cal G}\to0$, when aiming at the evaluation of 1PR diagrams by iteration of the operator~\eqref{eq:dFd} for the special case of ${\cal G}=0$, the leftmost operator can always be simplified to
\begin{equation}
 \frac{\partial {\cal L}^{1\text{-loop}}_\text{HE}}{\partial\bar F^{\mu\nu}}\frac{\partial}{\partial \bar F_{\mu\nu}}\bigg|_{{\cal G}=0}
 = \frac{1}{2} \frac{\partial {\cal L}^{1\text{-loop}}_\text{HE}({\cal F})}{\partial{\cal F}} \bar F_{\mu\nu}  \frac{\partial}{\partial \bar F_{\mu\nu}}\,.\label{eq:dFd|G=0}
\end{equation}

Apart from supplying an overall multiplying factor, the operator~\eqref{eq:dFd|G=0} effectively replaces an $\bar F$ in the function it acts on by itself.
Correspondingly, when acting with this operator on another factor of \Eqref{eq:dFd} and being ultimately interested in the limit of ${\cal G}=0$, the contribution $\sim{\cal G}$ in this factor can once again be dropped. 
This implies that when adopting the recursive procedure of Ref.~\cite{Karbstein:2017gsb} to construct higher-order 1PR contributions to ${\cal L}^{1\text{-loop}}_\text{HE}$
in constant fields fulfilling ${\cal G}=0$, all operators~\eqref{eq:dFd} can be replaced by \Eqref{eq:dFd|G=0} from the outset.

On the other hand, the rightmost of these operators effectively only acts on ${\cal L}^{1\text{-loop}}_\text{HE}({\cal F})$ and can thus be rewritten as
\begin{equation}
 \frac{\partial {\cal L}^{1\text{-loop}}_\text{HE}}{\partial\bar F^{\mu\nu}}\frac{\partial}{\partial \bar F_{\mu\nu}}\bigg|_{{\cal G}=0}
 = \frac{\partial {\cal L}^{1\text{-loop}}_\text{HE}({\cal F})}{\partial{\cal F}} {\cal F}  \frac{\partial}{\partial {\cal F}}\,.\label{eq:dFd|G=0_2ndline} 
\end{equation}
As \Eqref{eq:dFd|G=0_2ndline} only depends on $\cal F$, upon action of this operator we retain a function of $\cal F$, such that eventually all the operators \eqref{eq:dFd|G=0} can be substituted by \Eqref{eq:dFd|G=0_2ndline}.
In turn, for ${\cal G}=0$ the repeated action of the rather simple operator~\eqref{eq:dFd|G=0_2ndline} iteratively generates all the 1PR contributions to $\Gamma_{\rm HE}({\cal F})\big|_{N\to\infty}$ from ${\cal L}^{1\text{-loop}}_\text{HE}({\cal F})$.

Making use of the secular invariants of the electromagnetic field $c_\pm=(\sqrt{{\cal F}^2+{\cal G}^2}\pm{\cal F})^{1/2}$, for ${\cal G}=0$ and ${\cal F}>0$ (${\cal F}<0$) we have $c_+=\sqrt{2{\cal F}}$ and $c_-=0$ ($c_+=0$ and $c_-=\sqrt{-2{\cal F}}$).
Note that the case characterized by a finite value of $c_+$ ($c_-$) encompasses the scenario of a purely magnetic (electric) field. In line with this, we refer to it as magnetic-like (electric-like) field configuration.
Correspondingly, we finally obtain
\begin{equation}
 \frac{\partial {\cal L}^{1\text{-loop}}_\text{HE}}{\partial\bar F^{\mu\nu}}\frac{\partial}{\partial \bar F_{\mu\nu}}\bigg|_{{\cal G}=0}
 = \pm\frac{1}{2}\frac{\partial {\cal L}^{1\text{-loop}}_\text{HE}(c_\pm)}{\partial c_\pm} \frac{\partial}{\partial c_\pm}
 \label{eq:deriv+-}
\end{equation}
for $c_\mp=0$.
This identity implies that in the 't Hooft limit and a prescribed constant field fulfilling $c_\mp=0$ the 1PI tadpole is generated by $\partial {\cal L}^{1\text{-loop}}_\text{HE}(c_\pm)/\partial c_\pm$, and the effective coupling for $n$ tadpoles scales as $(\partial/\partial c_\pm)^n {\cal L}^{1\text{-loop}}_\text{HE}(c_\pm)$.
In fact, the above distinction of cases implies that a result derived for $c_+\neq0$, $c_-=0$ can be translated to the complementary case of $c_+=0$, $c_-\neq0$ by analytical continuation $c_+\to-{\rm i}c_-$. Correspondingly, it suffices to limit the explicit calculation to the case of $c_+\neq0$, $c_-=0$ only.

In this specific case, the $N$ flavor one-loop Heisenberg-Euler effective Lagrangian has the following closed-form representation \cite{Dittrich:1975au,Dittrich:1985yb,Heyl:1996dt,Dunne:2004nc},
\begin{align}
 {\cal L}_\text{HE}^{1\text{-loop}}(c_+)=N\frac{(ec_+)^2}{2\pi^2}\biggl\{&\zeta'\bigl(-1,\tfrac{1}{2}\tfrac{m^2}{ec_+}\bigr)+\frac{1}{4}\bigl(\tfrac{1}{2}\tfrac{m^2}{ec_+}\bigr)^2 \nonumber\\
 &-\frac{1}{2}\Bigl[\frac{1}{6}-\bigl(\tfrac{1}{2}\tfrac{m^2}{ec_+}\bigr)+\bigl(\tfrac{1}{2}\tfrac{m^2}{ec_+}\bigr)^2\Bigr]\ln\bigl(\tfrac{1}{2}\tfrac{m^2}{ec_+}\bigr)-\frac{1}{12}\biggr\} , \label{eq:L1loopG=0}
\end{align}
where $\zeta'(s,\chi)=\partial_s\zeta(s,\chi)$ denotes the first derivative of the Hurwitz zeta function.
The fact that the propertime integral has been performed explicitly in \Eqref{eq:L1loopG=0} indicates that -- at least in principle -- the contribution to $\Gamma_\text{HE}(c_\pm)|_{N\to\infty}$ at any loop order $\ell$ can be worked out in closed-form.
The strong field expansion of \Eqref{eq:L1loopG=0} follows from the series representations of the first derivative of the Hurwitz zeta function~\cite{Dowker:2015vya,Wolfram} and reads
\begin{align}
 {\cal L}_\text{HE}^{1\text{-loop}}(c_+)=N\frac{(ec_+)^2}{2\pi^2}&\biggl\{-\frac{1}{12}\ln\bigl(\tfrac{1}{2}\tfrac{m^2}{ec_+}\bigr) + \zeta'(-1)-\frac{1}{12} 
 -\frac{1}{2}\Bigl[\ln\bigl(\tfrac{1}{2}\tfrac{m^2}{ec_+}\bigr)-1+\ln(2\pi)\Bigr]\bigl(\tfrac{1}{2}\tfrac{m^2}{ec_+}\bigr) \nonumber\\
 -&\frac{1}{2}\Bigl[\ln\bigl(\tfrac{1}{2}\tfrac{m^2}{ec_+}\bigr)-\frac{3}{2}+\gamma\Bigr]\bigl(\tfrac{1}{2}\tfrac{m^2}{ec_+}\bigr)^2
 +\sum_{j=0}^\infty\frac{(-1)^j\zeta(j+2)}{(j+2)(j+3)}\bigl(\tfrac{1}{2}\tfrac{m^2}{ec_+}\bigr)^{j+3}\biggr\} , \label{eq:L1loopG=0sf}
\end{align}
where $\gamma$ is the Euler-Mascheroni constant, $\zeta(\cdot)$ is the Riemann zeta function and $\zeta'(\cdot)$ is its derivative.
Equation~\eqref{eq:L1loopG=0sf} immediately implies that the leading contribution to the $n$th derivative of ${\cal L}_\text{HE}^{1\text{-loop}}(c_+)$ with respect to the invariant $c_+$ in the strong field limit, $c_+\gg m^2/e$, scales as
\begin{equation}
 \frac{\partial^n{\cal L}_\text{HE}^{1\text{-loop}}(c_+)}{\partial c_+^n}\sim \biggl\{
  {c_+^{2-n}\ln c_+ \quad \text{for}\quad 0\leq n\leq2 \atop
  \;c_+^{2-n}\quad\text{for}\quad n\geq3 } \,.\label{eq:scalingwithBinsflimit}
\end{equation}

Therefore, in the strong field limit the 1PI tadpole scales as $\sim c_+\ln c_+$, and the effective two tadpole coupling as $\sim\ln c_+$.
On the other hand, effective couplings between $n\geq3$ tadpoles are parameterically suppressed with inverse powers of $c_+$.
Because of this, for $c_+\gg m^2/e$ and $c_-=0$ the dominant contribution to $\Gamma_\text{HE}|_{N\to\infty}$ at $\ell\geq2$ loops should stem from the diagram formed by two one-loop currents coupled by a chain of $\ell-2$ effective two-tadpole couplings.

The above excurse implies that aiming at the extraction of the strong field limit of $\Gamma_{\rm HE}|_{N\to\infty}$, all terms involving effective couplings of three or more tadpoles in \Eqref{eq:largeNGammafinal} can be dropped from the outset, and only the following contributions written out explicitly need to be considered
\begin{align}
 \Gamma_{\rm HE}[\bar A]\big|_{N\to\infty}=&-\frac{1}{4}\int\bar F_{\mu\nu}\bar F^{\mu\nu} + \Gamma_{\rm HE}^{1\text{-loop}}[\bar A]
 \nonumber\\
 &+\frac{1}{2}\iint \bigl(S^{(1)}_\psi[\bar A]\bigr)^\mu\bigl[\bigl(D^{-1}-S^{(2)}_\psi[\bar A]\bigr)^{-1}\bigr]_{\mu\nu} \bigl(S^{(1)}_\psi[\bar A]\bigr)^\nu\nonumber\\
 &+\text{terms involving effective couplings of $\geq3$ tadpoles}\,.
 \label{eq:largeNGammafinal_LOlargeB}
\end{align}

Upon limitation to a constant field characterized by $c_-=0$ and employing \Eqref{eq:Sn_constfield}, the terms written explicitly in \Eqref{eq:largeNGammafinal_LOlargeB} can be compactly expressed as
\begin{equation}
	 {\cal L}_{\rm HE}(c_+)\big|_{N\to\infty}=-\frac{1}{2}c_+^2 + {\cal L}_{\rm HE}^{1\text{-loop}}(c_+)
	 + \frac{1}{2}\,
 \frac{\frac{1}{2}\bigl(\frac{\partial {\cal L}_{\rm HE}^{1\text{-loop}}(c_+)}{\partial c_+}\bigr)^2}{1-\frac{1}{2}\bigl(\frac{1}{c_+}\frac{\partial {\cal L}_{\rm HE}^{1\text{-loop}}(c_+)}{\partial c_+}+\frac{\partial^2 {\cal L}_{\rm HE}^{1\text{-loop}}(c_+)}{\partial c_+^2}\bigr)} \,. \label{eq:strongBallloop}
\end{equation}
Accounting only for the leading contribution to \Eqref{eq:L1loopG=0} in the strong field limit, $ec_+/m^2\gg1$, i.e., ${\cal L}_{\rm HE}^{1\text{-loop}}(c_+)\simeq\frac{N}{24\pi^2}(ec_+)^2\ln(\frac{ec_+}{m^2})$, \Eqref{eq:strongBallloop} becomes \cite{Karbstein:2019wmj}
\begin{equation}
	 {\cal L}_{\rm HE}(c_+)\big|_{N\to\infty}\simeq-\frac{1}{2}c_+^2 + \frac{N}{8\pi}\frac{1}{3\pi}(ec_+)^2\ln\Bigl(\frac{ec_+}{m^2}\Bigr)\biggl[1+\frac{1}{2}\,
 \frac{N\frac{\alpha}{3\pi}\ln(\frac{ec_+}{m^2})}{1-N\frac{\alpha}{3\pi}\ln(\frac{ec_+}{m^2})}\biggr] \,.
 \label{eq:sfL_c+}
\end{equation}

Two comments are in order in this context: first, we note that the approach employed here directly provides the result for the all-order result derived in  Ref.~\cite{Karbstein:2019wmj} without the need of working out any combinatorial symmetry factors.
Second, we emphasize that as it is manifestly based on a perturbative expansion in the number of loops this result for the all-loop strong field limit, $ec_+/m^2\gg1$, a priori only allows for trustworthy insights into parameter regimes where a loop expansion is possible and makes sense, i.e., $N\frac{\alpha}{3\pi}\ln(\frac{ec_+}{m^2})<1$.
The corresponding fully non-perturbative result not relying on a loop expansion could be extracted along the lines outlined below \Eqref{eq:barjconst}.

On the other hand, for electric-like field configurations characterized by $c_-\neq0$ and $c_+=0$ the result of the Heisenberg-Euler Lagrangian in the strong field limit, $ec_-/m^2\gg1$,  follows from \Eqref{eq:sfL_c+} upon substitution of $c_+\to{\rm e}^{-{\rm i}\frac{\pi}{2}}c_-$; cf. the discussion below Eq.~\eqref{eq:deriv+-}.
As to be expected, ${\cal L}_{\rm HE}$ develops an imaginary part in electric-like field configuration. The explicit expressions for the real and imaginary parts constituting its strong field limit read \cite{Karbstein:2019wmj}
\begin{align}
	 \Re\{{\cal L}_{\rm HE}(c_-)\}\big|_{N\to\infty}&\simeq\frac{1}{2}c_-^2 - \frac{N}{8\pi}\frac{1}{3\pi}(ec_-)^2\ln\Bigl(\frac{ec_-}{m^2}\Bigr)\biggl[1+\frac{1}{2}\,
 \frac{N\frac{\alpha}{3\pi}\ln(\frac{ec_-}{m^2})}{1-N\frac{\alpha}{3\pi}\ln(\frac{ec_-}{m^2})}\biggr] \,, \nonumber\\
	 \Im\{{\cal L}_{\rm HE}(c_-)\}\big|_{N\to\infty}&\simeq  \frac{N}{8\pi}\frac{1}{6}(ec_-)^2\biggl[1+
 \frac{N\frac{\alpha}{3\pi} \ln(\frac{ec_-}{m^2})}{1-N\frac{\alpha}{3\pi}\ln(\frac{ec_-}{m^2})}\biggr] \,.
 \label{eq:sfL_c-}
\end{align}
As a side remark, it is interesting to note that in the formal limit of $N\frac{\alpha}{3\pi}\ln(\frac{ec_+}{m^2})\gg1$, i.e., clearly beyond the validity regime of perturbation theory and the Landau pole, the factor of unity in the denominator can be neglected and thus in particular $\Im\{{\cal L}_{\rm HE}(c_-)\}\big|_{N\to\infty}\to0$.

Finally, we emphasize that upon setting $N=1$ in Eqs.~\eqref{eq:sfL_c+} and \eqref{eq:sfL_c-} the correct strong field limits of standard single-flavor QED are recovered even though standard QED is certainly not a large $N$ theory. The reason for this is that the dominant strong field behavior of the Heisenberg-Euler effective action for standard QED is precisely encoded in the bubble chain diagrams persisting in the large $N$ limit; cf. also the corresponding discussion in Ref.~\cite{Karbstein:2019wmj} and references therein.

\section{Conclusions and Outlook}\label{sec:concls_N}

In this article, we studied external-field QED with $N$ charged particle flavors, putting special attention on the large $N$ limit. Similar considerations are, of course, also possible for scalar QED.
A central and important result of the present study are closed-form expressions for the Heisenberg-Euler effective action for this theory in constant fields.
These findings allow for new insights into the strong-field limit of the theory.

We are convinced that external-field QED in the large $N$ limit constitutes a very interesting deformation of standard external-field QED, and believe that its study will also have large feedback and relevance for the latter.
The present work in particular lays the foundations for a future non-perturbative first-principles study of the large $N$ Heisenberg-Euler effective action in arbitrarily strong constant electromagnetic fields.

\acknowledgments

This work has been funded by the Deutsche Forschungsgemeinschaft (DFG) under Grant No. 416607684 within the Research Unit FOR2783/1.

\end{document}